\documentclass[%
 aip,cha,amsmath,amssymb,numerical,
preprint,%
]{revtex4-1}
\usepackage{makecell}
\usepackage{graphicx}
\usepackage{dcolumn}
\usepackage{bm}
\usepackage{mathtools} 

\renewcommand{\th}{%
    \ifmmode
        ^\mathrm{th}%
    \else%
        \textsuperscript{th}\xspace%
    \fi%
}

\begin{document}

\preprint{AIP/123-QED}

\title{Multitarget search on complex networks: A logarithmic growth of global mean random cover time}
\author{Tongfeng Weng}
\email{wtongfeng2006@163.com}
 \affiliation{HKUST-DT System and Media Laboratory, Hong Kong University of Science and Technology, HongKong}
 
 \author{Jie Zhang}
 \affiliation{Centre for Computational Systems Biology, Fudan University, China}
 \author{Michael Small}
 \affiliation{The University of Western Australia, Crawley, WA 6009, Australia}
 \affiliation{Mineral Resources, CSIRO, Kensington, WA, Australia}
 \author{Ji Yang}
 \affiliation{HKUST-DT System and Media Laboratory, Hong Kong University of Science and Technology, HongKong}
 \author{Farshid Hassani Bijarbooneh}
 \affiliation{HKUST-DT System and Media Laboratory, Hong Kong University of Science and Technology, HongKong}
\author{Pan Hui}%
 \email{panhui@cse.ust.hk}
\affiliation{HKUST-DT System and Media Laboratory, Hong Kong University of Science and Technology, HongKong}%

\
\date{\today}

\begin{abstract}
We investigate multitarget search on complex networks and derive an exact expression for the mean random cover time that quantifies the expected time a walker needs to visit multiple targets. Based on this, we recover and extend some interesting results of multitarget search on networks. Specifically, we observe the logarithmic increase of the global mean random cover time with the target number for a broad range of random search processes, including generic random walks, biased random walks, and maximal entropy random walks. We show that the logarithmic growth pattern is a universal feature of multi-target search on networks by using the annealed network approach and the Sherman-Morrison formula. Moreover, we find that for biased random walks, the global mean random cover time can be minimized, and that the corresponding optimal parameter also minimizes the global mean first passage time, pointing towards its robustness. Our findings further confirm that the logarithmic growth pattern is a universal law governing multitarget search in confined media.
\end{abstract}

\maketitle

\begin{quotation}
It has been recognized that random search processes are an important branch of network science. The importance originates from their broad relevance ranging from diseases spreading, animal foraging, to biochemical reactions. Previous studies of random search processes mainly concentrated on the discovery of a single target, while much less is known about the search time of finding more than one target given in advance. In this paper, we investigate multitarget search on complex networks and propose an iterative approach to derive mean random search time analytically. We show that the growth of mean random search time at a global scale seems to follow a logarithmic function of the number of targets. Furthermore, we find evidence that this logarithmic growth pattern is a universal principle governing multi-object search across various random search strategies including generic random walks, biased random walks, and maximal entropy random walks.
\end{quotation}

\section{Introduction}

Random search processes have attracted increasing investigation over the past decade \cite{JDNoh2004,MStarnini2012,JHPeng2015,TFWeng2016}, due to their broad relevance to various circumstances ranging from diseases and information spreading \cite{ALLloyd2001}, animal foraging \cite{Viswanathan2011,VPalyulin2014}, to transport in media \cite{DBenAvraham2000}. So far, most studies of random searches have been limited to single target discovery \cite{JDNoh2004,MStarnini2012,JHPeng2015,TFWeng2016}. However, in the information age, multiple targets usually need to be found simultaneously, a problem which is commonly encountered in the fields of chemistry, biology and social interaction. Examples range from immune-system cells chasing pathogens \cite{LMHeuze2013}, robotic task allocation \cite{MVergassola2007}, to animals foraging \cite{Viswanathan2011}. In fact, the trapping problem of multiple targets has already received great attention \cite{HScher1981,LKGallos2004,EAgliari2010,BMeyer2012,EAgliari2007}. Extensive works have been devoted to evaluating this trapping problem, such as a concentration of static traps on scale-free networks \cite{LKGallos2004} or on recursive networks \cite{BMeyer2012} and even a number of mobile traps on low-dimensional substrates \cite{EAgliari2007}. Going beyond the trapping aspect, another desirable quantity for characterizing multi-object search is the mean random cover time, which quantifies the expected time needed to find several sites specified in advance. Characterization of this quantity has been a long-standing problem in the realm of random walk theory due to its broad relevance \cite{Dembo2004,JMendonca2011}. 

However, studies of mean random cover time remain scarce and are still in the early stage. Nemirovsky $et$ $al.$ reveal the universality of cover time on regular cubic lattices \cite{Nemirovsky1990} --- that is the extreme case where all sites of a given domain need to be visited. Later, Coutinho $et$ $al.$ analyze mean random cover time in two dimensions using Monte Carlo simulations \cite{KRCoutinho1994}. Recently, Nascimento $et$ $al.$ provide some analytical results of mean random cover time in one dimensional lattices \cite{MSNascimento2001}. In fact, most studies either focus on the problems of mean random cover time or cover time on regular graphs  \cite{Nemirovsky1990,Dembo2004,MSNascimento2001,JMendonca2011} or provide numerical results of the random cover time \cite{KRCoutinho1994}. Very recently, Chupeau $et$ $al.$ reveal the universal form of the full distribution of the partial and random cover time \cite{MChupeau2015}, which makes an important step in multiple targets search. Interestingly, the first moment of the random cover time seems to imply a logarithmic growth pattern of the search time versus the target number. Nonetheless, a general framework for mean random cover time that allows one to calculate this analytically on an arbitrary network has not yet been constructed.

In this paper, we study the multi-target search on diverse networks and propose an iterative approach to determine the mean random cover time (MRCT) of complex networks analytically. The quantity MRCT quantifies the expected time required for a searcher to find a number of targets given in advance. Based on this analytical derivation, we find the slow (logarithmic) increase of the global MRCT with the target number, which is much smaller than the linear growth one intuitively expects. Remarkably, we show that this relationship is a universal principle governing multi-object search for various random search processes including generic random walks, biased random walks, and maximal entropy random walks. Our findings further enrich our understanding of multitarget search in nature. 

This remainder of this paper is organized as follow: In Sec. \ref{sec1}, we provide an iterative approach to derive the explicit expression of mean random cover time. This approach is applied to generic random walks described in Section \ref{sec2}. In Sec \ref{sec3} and Sec \ref{sec4}, we analyze multitarget search of a biased random walk strategy and maximal entropy random walk strategy, respectively. Our conclusion is given in Sec. \ref{sec5}.

\section{Explicit expression for mean random cover time}\label{sec1}
\begin{figure}[!htb]
\centering
\includegraphics[width=0.4\textwidth,height=0.27\textheight]{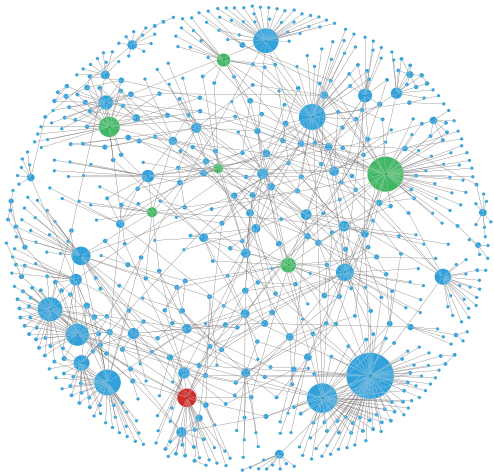}
\caption{(Color online) An example of multi-object search on the ``Yeast'' network \cite{HJeong2001}: the expected time needed to find six distinct nodes (colored green) for a walker starting from the source node (colored red) with no prior knowledge of target distribution, defined as the mean random cover time.}\label{f1}
\end{figure}

We consider a random walker traveling on a network consisting of $N$ nodes. The connectivity is represented by the adjacency matrix A, whose entries $a_{ij}=1$ (or 0) if there is (not) a link from nodes $i$ to $j$. At each time step, the walker moves from current node $i$ to node $j$ with the transition probability $p_{ij}$, which constitutes the $ij\th$ entry of transition matrix P. Take generic random walks for example, the transition probability is $p_{ij}=a_{ij}/k_{i}$, where $k_{i}=\sum_{j}a_{ij}$ is the degree of node $i$. Here, we are interested in how long does it take the walker to reach several target nodes for the first time, known as the MRCT $T^{(m)}_{i,\Omega_{m}}$ --- the expected time needed to visit $m$ distinct nodes $\Omega_{m}=\{v_{1},v_{2},\cdots,v_{m}\}$ starting from node $i$ (see Fig.~\ref{f1}). In particular, when $m=1$, the mean random cover time reduces to the mean first passage time, to which many previous studies have been devoted \cite{JDNoh2004}. To derive the MRCT analytically, we first consider a simple case of two target search and assume that the two targets are placed at nodes $v_{1}$ and $v_{2}$. In this situation, if the first step of the walker is to node $v_{1}$ (resp. $v_{2}$), the expected number of steps required is $T_{v_{1},v_{2}}+1$ (resp. $T_{v_{2},v_{1}}+1$); if it is to some other node $j$, the expected number of steps becomes $T_{j,\{v_{1},v_{2}\}}+1$. Thus, for $i\neq{v_{1}}$ and ${v_{2}}$, we have
\begin{equation}
\ T_{i,\{v_{1},v_{2}\}}=p_{iv_{1}}(T_{v_{1},v_{2}}+1)+p_{iv_{2}}(T_{v_{2},v_{1}}+1)+\sum_{j\neq{v_{1},v_{2}}}p_{ij}(T_{j,\{v_{1},v_{2}\}}+1).
\label{e1}
\end{equation}
From Eq.~(\ref{e1}), we can express the MRCT $T_{i,\{v_{1},v_{2}\}}$ in terms of the associated mean first passage time analytically as follows (see Appendix)
\begin{equation}
\ T_{i,\{v_1,v_2\}}=\frac{T_{v_1,v_2}T_{v_2,v_1}+T_{i,v_1}T_{v_2,v_1}+T_{i,v_2}T_{v_1,v_2}}{T_{v_1,v_2}+T_{v_2,v_1}}.
\label{s14}
\end{equation}
Repeatedly, suppose that we have already obtained the MRCT $T^{(m-1)}_{i,\Omega_{m-1}}$ for $m-1$ targets search on the network. Consequently, we will consider how to derive the MRCT $T^{(m)}_{i,\Omega_{m}}$ exactly from the known MRCT. Similarly, it is easy to verify that the equations $T^{(m)}_{i,\Omega_{m}}=T^{(m-1)}_{i,(\Omega_{m}{\setminus}i)}$ hold for $i\in{\{v_1,v_2,\cdots,v_m\}}$. Regarding $i\notin{\Omega_{m}}$, we have 
\begin{equation}
\ T^{(m)}_{i,\Omega_{m}}=\sum_{v_{j}\in{\Omega_{m}}}p_{iv_{j}}(T^{(m-1)}_{v_{j},(\Omega_{m}{\setminus}v_{j})}+1)+\sum_{l\notin{\Omega_{m}}}p_{il}(T^{(m)}_{l,\Omega_{m}}+1).
\label{e3}
\end{equation}
We can rewrite Eq.~(\ref{e3}) in matrix form as
\begin{equation}
\ T^{(m)}_{\Omega_{m}}=\bar{\textbf{e}}+\sum_{v_{j}\in{\Omega_{m}}}\bar{P}_{v_{j}}\times{T_{v_{j},(\Omega_{m}{\setminus}v_{j})}\bar{\textbf{e}}}+\bar{P}T^{(m)}_{\Omega_{m}},
\label{e4}
\end{equation}
where $T^{(m)}_{\Omega_{m}}$ is an $(N-m)$-dimensional vector $(T^{(m)}_{i,\Omega_{m}}\mid{i\notin{\Omega_{m}}})$; $\bar{\textbf{e}}$ is the all-ones vector; $\bar{P}$ is the submatrix of the transition probability matrix P obtained by deleting the set of rows and columns with indexes $\{v_{i}\mid{v_{i}\in{\Omega_{m}}}\}$; $\bar{P}_{v_{i}}$ represents the $v_{i}\th$ column of the matrix $P$ without the elements $\{p_{v_{j},v_{i}}\mid{v_{j}\in{\Omega_{m}}}\}$. Since the matrix $(I-\bar{P})$ is reversible \cite{Grinstead2006}, we have
\begin{equation}
\ T^{(m)}_{\Omega_{m}}=(I-\bar{P})^{-1}\left(\bar{\textbf{e}}+\sum_{v_{j}\in{\Omega_{m}}}\bar{P}_{v_{j}}\times{T^{(m-1)}_{v_{j},(\Omega_{m}{\setminus}v_{j})}\bar{\textbf{e}}}\right).
\label{e5}
\end{equation}
Equation (\ref{e5}) is important as it provides a universal principle for calculating the MRCT iteratively. More importantly, this expression allows us to link the gap between mean first passage time $(m=1)$ to cover time $(m=N-1)$, and thereby to probe the intermediate region $1<{m}<N-1$, about which little is known. Note that it is  theoretically possible to express the MRCT $T^{(m)}_{i,\Omega_{m}}$ in terms of the mean first passage time resembling Eq.~(\ref{s14}), which can benefit us for computing $T^{(m)}_{i,\Omega_{m}}$ directly. Unfortunately, the expression will become rather lengthy and does not seem to be practical in the situation where $m$ is large. Nonetheless, our iterative approach, for the first time, provides an useful way of calculating mean random cover time analytically on an arbitrary network. 

\begin{figure}[!htb]
\centering
\includegraphics[width=0.95\textwidth,height=0.27\textheight]{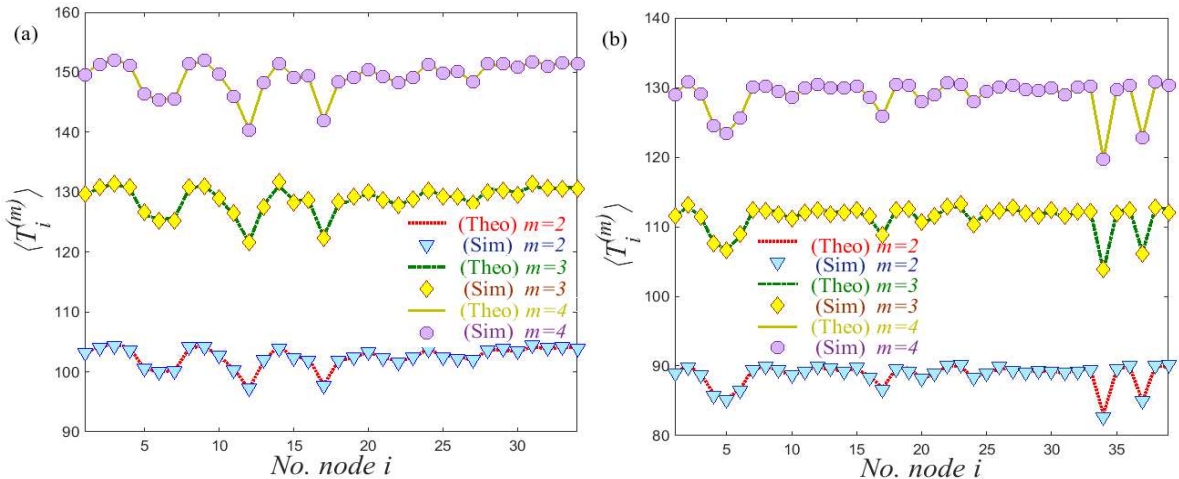}
\caption{(Color online) The effort of source node on multiple targets search for (a) the ``karate club'' network \cite{Zachary1977} and (b) the ``Chesapeake'' network \cite{BairdD1989}. All data collapse to the theoretical results given by Eq.~(\ref{e5}).}\label{f2}
\end{figure}

We now confirm the analytical results by Monte Carlo simulations for generic random walks taking place in the ``karate club'' network \cite{Zachary1977} and the ``Chesapeake'' network \cite{BairdD1989}. To achieve the numerical results, we compute the time required for a walker to travel from a source node to multiple target nodes given in advance and average over the ensemble of 50,000 independent runs. Figure~\ref{f2} shows an excellent agreement between the analytical results and the numerical simulations. The prediction of Eq.~(\ref{e5}) unambiguously captures the time required to find multiple targets, as expected. Meanwhile, we notice that the profiles of the quantity $\langle{T^{(m)}_{i}}\rangle=\frac{1}{\binom{N-1}{m}}\sum_{\Omega_{m}}T_{i,\Omega_{m}}$ --- characterizing the effects of source location on multi-object search, present the same tendency with respect to source position for different number of targets $m$. These results indicate that the effects of source site seem to be independent of the number of targets. 

\section{The logarithmic growth pattern of generic random walks}\label{sec2}
\begin{figure}[!htb]
\centering
\includegraphics[width=0.95\textwidth,height=0.75\textheight]{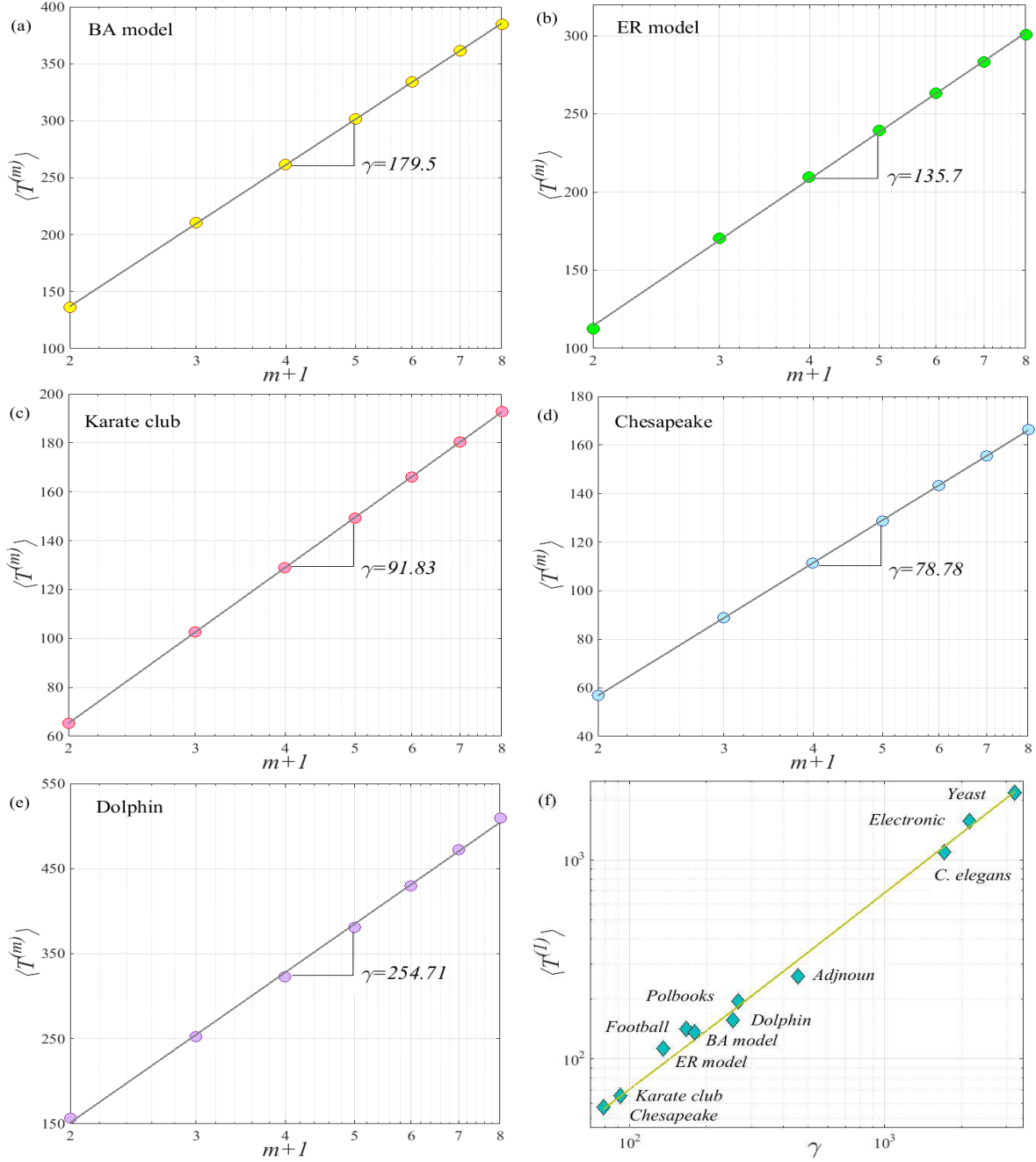}
\caption{(Color online) The semilogarithmic plots show the global MRCT $\langle{T^{(m)}}\rangle$ as a function of number of targets $m$ on (a) the BA model, (b) the ER model, (c) the ``Karate club'' network, (d) the ``Chesapeake'' network, and (e) the ``Dolphin'' network. The values of $\gamma$ are obtained from the slopes of the fitting straight lines. (f) Relationship between the growth rate $\gamma$ and mean first passage time $\langle{T^{(1)}}\rangle$ on a number of considering synthetic and real networks. The straight line refers to the best fit $\langle{T^{(1)}}\rangle=0.68\gamma-2.2$.}\label{f3}
\end{figure}

In practice, one is usually more concerned with how the mean random cover time increases with the target number as it dictates how long one will need to reach a new target. Here, to evaluate search time at a global scale, we introduce the global MRCT $\langle{T^{(m)}}\rangle$ defined by
\begin{equation}
\ \langle{T^{(m)}\rangle}=\frac{1}{N\binom{N-1}{m}}\sum_{i}\sum_{\Omega_{m}}T^{(m)}_{i,\Omega_{m}}.
\label{s20}
\end{equation}
We investigate the global MRCT $\langle{T^{(m)}}\rangle$ as a function of target number $m$ for two synthetic networks (the Barab\'{a}si-Albert (BA) model \cite{ABarabasi1999} and the Erd\"{o}s-R\'{e}nyi (ER) model \cite{PErdos1950}) and three real networks (the ``Karate club'' network \cite{Zachary1977}, the ``Chesapeake'' network \cite{BairdD1989}, and the ``Dolphin'' network \cite{DLusseau2003}). Interestingly, the results of Fig.~{\ref{f3}} show that the way in which $\langle{T^{(m)}}\rangle$ scales with $m$ seems to follow a logarithmic behavior such that $\langle{T^{(m)}}\rangle\approx{\gamma{ln(m+1)}}$, where $\gamma$ represents the growth rate of search time. This growth pattern is much smaller than the linear relationship which one would intuitively expect. This suggests that much less time is needed to find an extra new target in a multiple targets search. Utilizing the annealed network approach \cite{SNDorogovtsev2008} and the Sherman-Morrison formula \cite{JSherman1950}, we present analytical arguments to explain the universal growth pattern of $\langle{T^{(m)}}\rangle$ versus $m$. For an uncorrelated network, we can reinterpret the adjacency matrix $A$ as a weighted fully connected graph $\widetilde{A}$ based on the annealed network approach \cite{SNDorogovtsev2008}. Specifically, the entry $\widetilde{a}_{ij}=\frac{k_{i}k_{j}}{N\langle{k}\rangle}$ defines the connection probability between nodes $i$ and $j$, where $\langle{k}\rangle$ represents the average degree of the whole network. In this situation, the transition probability of the generic random walks becomes
\begin{equation}
\ P=\frac{1}{\sum_{i=1}^{N}k_{i}}\textbf{e}(k_{1},k_{2},\cdots,k_{N}),
\label{s17}
\end{equation}
where $\textbf{e}$ is a $N$-dimensional column vector with all entries $1$. Utilizing the Sherman-Morrison formula \cite{JSherman1950}, the inverse of the matrix $(I-\bar{P})$ becomes
\begin{equation}
\  (I-\bar{P})^{-1}=\frac{1}{\sum_{i=1}^{m}k_{v_{i}}}\left(\begin{array}{cccc}
 k_{1}+\sum_{i=1}^{m}k_{v_{i}}&k_{2}& \cdots &k_{N}\\
k_{1}&k_{2}+\sum_{i=1}^{m}k_{v_{i}}& \cdots &k_{N}\\
 \vdots&\vdots&\vdots&\vdots\\
k_{1}&k_{2}& \cdots &k_{N}+\sum_{i=1}^{m}k_{v_{i}}
 \end{array} \right)_{(N-m)\times(N-m)}.
\label{s18}
\end{equation}
Inserting Eq.~(\ref{s18}) into Eq.~(\ref{e5}) with a few simple algebraic manipulations, we obtain
\begin{equation}
\ T^{(m)}_{\Omega_{m}}=\left(\frac{N\langle{k}\rangle}{\sum_{i=1}^{m}k_{v_{i}}}+\frac{\sum_{i=1}^{m}k_{v_{i}}T^{(m-1)}_{v_{i},\Omega_{m}{\setminus}v_{i}}}{\sum_{i=1}^{m}k_{v_{i}}}\right)\times{\bar{\textbf{e}}}.
\label{s19}
\end{equation}
Substitution into Eq.~(\ref{s20}) gives
\begin{equation}
\ \langle{T^{(m)}\rangle}\approx\frac{N}{m}+\langle{T^{(m-1)}\rangle}.
\label{s21}
\end{equation}
Thus, we have a recursion relation for the global MRCT for $m$ targets in terms of $m-1$ targets. In this situation, since it is easy to verify that $\langle{T^{(1)}}\rangle\approx{N}$, equation~(\ref{s21}) can be solved to obtain
\begin{equation}
\ \langle{T^{(m)}}\rangle\approx{\langle{T^{(1)}}\rangle{\sum_{i=1}^{m}\frac{1}{i}}}.
\label{e6}
\end{equation}
Using the lower bound $ln(m+1)$ for estimating the partial sums of the harmonic series $\sum_{i=1}^{m}1/i$, we have 
 \begin{equation}
\ \langle{T^{(m)}}\rangle\approx{\gamma}ln(m+1),
\label{es}
\end{equation}
where $\gamma$ represents the growth rate. In particular, when $m=1$, we have $\langle{T^{(1)}}\rangle\approx{ln(2)\gamma}$, which hints at an approximate value of the growth rate $\gamma$. Figure \ref{f3}(f) further supports the validity of this approximation by showing a linear relationship between $\langle{T^{(1)}}\rangle$ and $\gamma$ (i.e., $\langle{T^{(1)}}\rangle=0.68\gamma-2.2$) on synthetic and real networks (as shown in Table~\ref{t1}), which is consistent with our theoretical prediction (i.e., $\langle{T^{(1)}}\rangle\approx{ln(2)\gamma}$). The result of Eq.~(\ref{es}) reveals that the growth of the global MRCT follows a logarithmic pattern for multi-object search in nature.

\begin{table}
\caption{Summary of the details of the real-world networks. For each network, its size $N$, the number of links $E$, the average path length $\langle{d}\rangle$, the assortativity coefficient $r$, the global MFPT $\langle{T^{(1)}}\rangle$, the growth rate $\gamma$, and the description of the network, are given.}
\begin{tabular}{ccccccccccccccc}
\hline
\hline
Data Sets&&N&&E&&$\langle{d}\rangle$&&$r$&&$\langle{T^{(1)}}\rangle$&&$\gamma$&&Description\\
\hline
Yeast \cite{HJeong2001}&&662&&1062&&5.20&&-0.41&&2186.6&&3241.8&&\makecell[c]{Network of regulatory proteins and \\genes in the yeast S. cerevisiae}\\
\hline
Karate club \cite{Zachary1977}&&34&&78&&2.41&&-0.47&&65.38&&91.83&&\makecell[c]{Social network of friendships \\within a karate club}\\
\hline
Chesapeake \cite{BairdD1989}&&39&&170&&1.83&&-0.37&&56.96&&78.78&&Chesapeake bay mesohaline network\\
\hline
Adjnoun \cite{MEJNewan2006}&&112&&425&&2.53&&-0.13&&259.4&&458.5&&\makecell[c]{Adjacency network of common \\adjectives and nouns} \\
\hline
Electronic \cite{RMilo2002}&&512&&819&&6.86&&-0.03&&1574.5&&2155.1&&\makecell[c]{Adjacency network of electronic \\sequential logic circuits}\\
\hline
Dolphin \cite{DLusseau2003}&&62&&159&&3.36&&-0.04&&156.8&&254.7&&\makecell[c]{Network of dolphins in a community \\ living in Doubtful Sound}\\
\hline
Football \cite{MGirvan2002}&&115&&615&&2.51&&0.16&&141.1&&166.7&&American college football\\
\hline
C. elegans \cite{HJeong2000}&&453&&2025&&2.66&&-0.22&&1098.5&&1713.1&&Metabolic network of C. elegans\\
\hline
Polbooks \cite{MRipeanu2002}&&105&&441&&3.08&&-0.13&&194.9&&266.4&&\makecell[c]{Network of books on USA \\politics around 2004}\\
\hline
\end{tabular}
\label{t1}
\end{table}

\section{Global mean random cover time of biased random walks}\label{sec3}
\subsection{The effect of the tuning parameter on global mean random cover time}
\begin{figure}[!htb]
\centering
\includegraphics[width=0.95\textwidth,height=0.53\textheight]{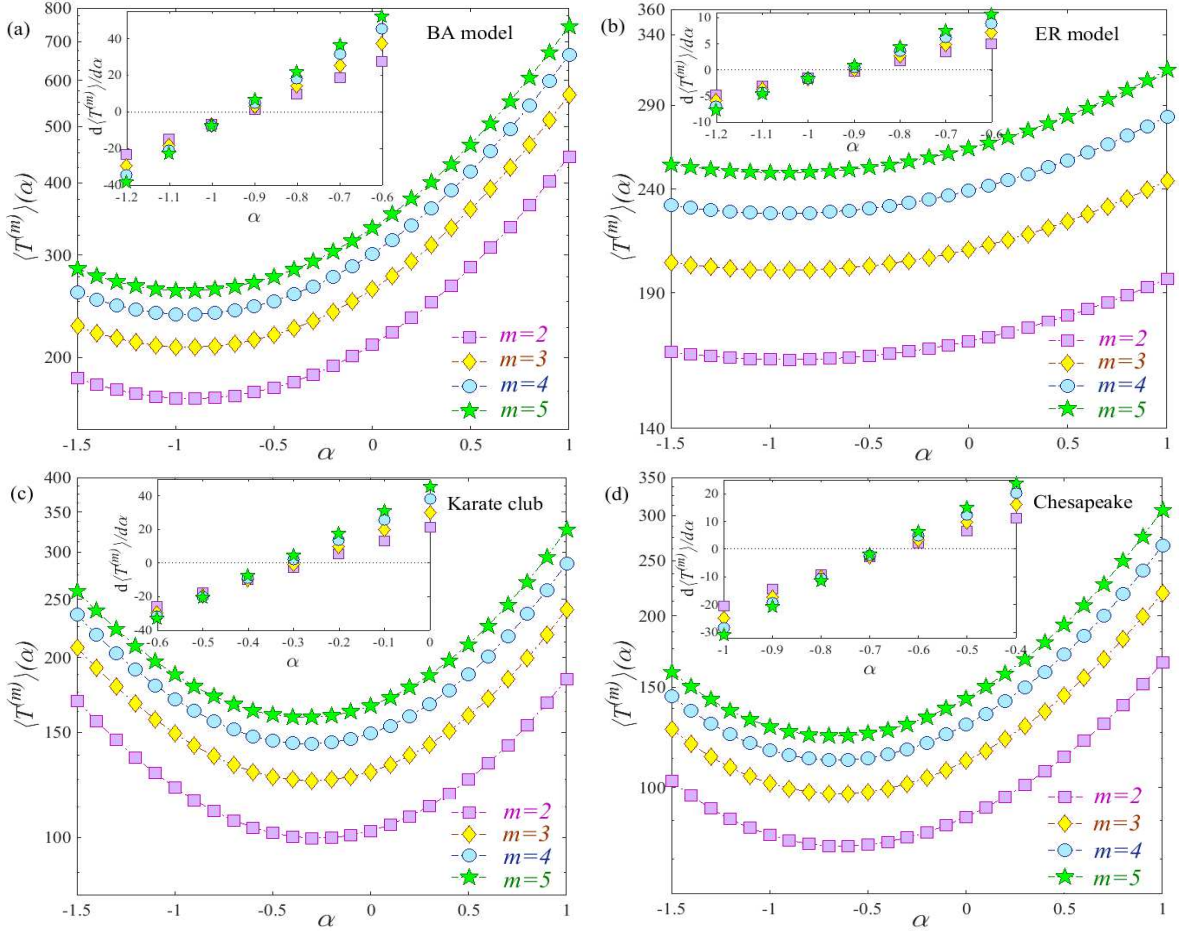}
\caption {(Color online) The global MRCT as a function of $\alpha$ over different number of targets $m$ on (a) the BA model, (b) the ER model, (c) the ``Karate club'' network, and (d) the Chesapeake'' network. The insets show the first derivative $d\it{\langle{T^{(m)}}\rangle}/d\alpha$ versus the tuning parameter $\alpha$.} \label{fig2}
\end{figure}

As a further validation of the logarithmic growth pattern, we address the general case of biased random walks on various networks. Specifically, at each time step, the walker moves from current node $i$ to node $j$ with transition probability $p_{ij}=\frac{a_{ij}k_j^\alpha}{\sum_ja_{ij}k_j^\alpha}$, where $\alpha$ is the tuning parameter \cite{AFronczak2009}. Clearly, the tuning parameter $\alpha$ controls the preference of visiting high or low degree node in each time step, which in turn fully determines the behaviors of the biased random walks. To quantify the search efficiency of a biased random walker with respect to the tuning exponent $\alpha$, we obverse the behavior of $\langle{T^{(m)}}\rangle$ versus $\alpha$ on various networks including two synthetic networks (the BA model \cite{ABarabasi1999} and the ER model \cite{PErdos1950})  and two real networks (the ``Karate club'' network \cite{Zachary1977} and the ``Chesapeake'' network \cite{BairdD1989}), as shown in Fig.~\ref{fig2}. Clearly, for each network, all profiles present the same tendency with increasing number of targets. In particular, the results presented in Fig.~\ref{fig2} clearly show the presence of a minimum $\langle{T^{(m)}}\rangle$ for different $m$ at the same position. This is further supported by observing the first derivative $d\langle{T^{(m)}}\rangle/{d\alpha}$ versus the tuning parameter $\alpha$, where the optimal tuning exponent $\alpha_{\rm{opt}}$ (i.e., $d\langle{T^{(m)}}\rangle/{d\alpha}$ nears zeros.) occurs at the same point for each network as illustrated in the insets of Fig.~\ref{fig2}. These results hint that an optimal tuning exponent $\alpha_{\rm{opt}}$ of biased random walks is independent of number of targets $m$. This finding is consistent with the results reported in Ref. 21. On the other hand, the results point out that to reach the efficient mobility of multi-target search for biased random walks, we can adopt the strategy as that used for finding the optimal tuning exponent $\alpha_{\rm{opt}}$ in one target search \cite{MBonaventura2014}. In particular, from Fig.~\ref{fig2} (a) and (b), we can see that $\alpha_{\rm{opt}}\approx{-1}$ for the BA and ER models. These findings are consistent with the results of one target search reported in Ref. 37, where for an uncorrelated network, the optimal tuning parameter is $\alpha_{\rm{opt}}\approx{-1}$.

\subsection{The logarithmic growth pattern of biased random walks}
\begin{figure}[!htb]
\centering
\includegraphics[width=0.95\textwidth,height=0.53\textheight]{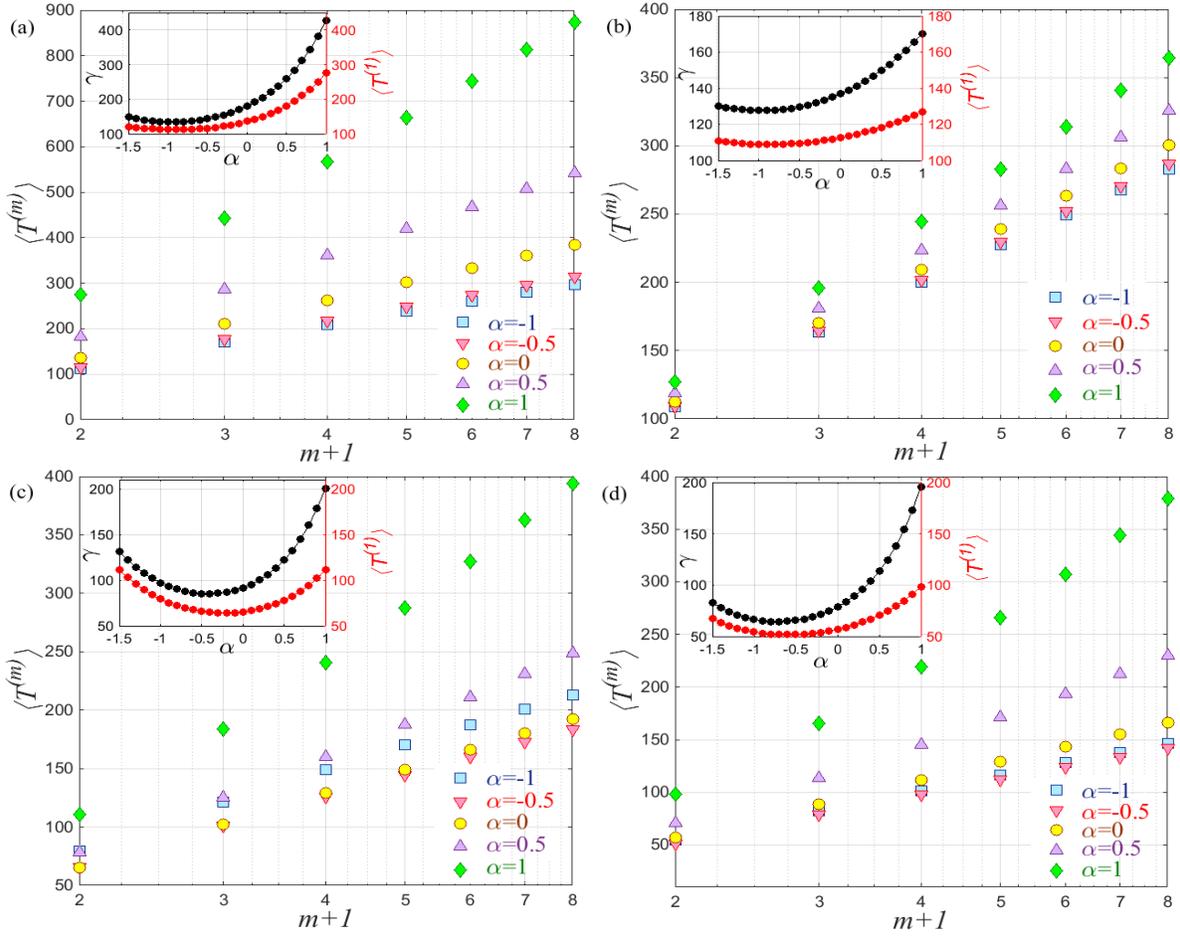}
\caption {(Color online) The semilogarithmic plots show the global MRCT $\langle{T^{(m)}}\rangle$ as a function of $m$ over different tuning exponents $\alpha$ on (a) the BA model, (b) the ER model, (c) the ``Karate club'' network, and (d) the Chesapeake'' network. In the insets, we show the profiles of the estimated values of $\gamma$ as a function of $\alpha$ in comparison with that of $\langle{T^{(1)}}\rangle$ vs $\alpha$.} \label{fig3}
\end{figure}

Moreover, we investigate the global MRCT $\langle{T^{(m)}}\rangle$ as a function of number of targets $m$ for the biased random walks with respect to different tuning parameters $\alpha$ on the previously considering networks. Interestingly, Figure~{\ref{fig3}} shows that the way in which $\langle{T^{(m)}}\rangle$ scales with $m$ seems to follow the logarithmic behavior such that $\langle{T^{(m)}}\rangle\varpropto{{ln(m+1)}}$. The results further demonstrate that the logarithmic growth mechanism is a universal principle governing multiple target search. Utilizing the annealed network approach  \cite{SNDorogovtsev2008} and the Sherman-Morrison formula \cite{JSherman1950}, we theoretically explain why this interesting growth pattern emerges even for biased random walks. In the same manner, we first reinterpret the adjacency matrix $A$ as a weighted fully connected graph $\widetilde{A}$ based on the annealed network approach \cite{SNDorogovtsev2008}. In this situation, the transition probability $P$ of biased random walks becomes
\begin{equation}
\ P=\frac{1}{\sum_{i=1}^{N}k_{i}^{1+\alpha}}\textbf{e}( k_{1}^{1+\alpha},k_{2}^{1+\alpha},\cdots,k_{N}^{1+\alpha}).
\label{e9}
\end{equation}
Repeating similar calculations as we did for the previous random walks, we obtain the identical result given already by Eq.~(\ref{e6}). Moreover, the strong correlation between $\gamma$ and $\langle{T^{(1)}}\rangle$ is further supported by observing their behaviors as a function of $\alpha$ as illustrated in the insets of Fig.~\ref{fig3}, where the profile of $\gamma$ versus $\alpha$ present the same tendency as that of $\langle{T^{(1)}}\rangle$ vs $\alpha$ on each network. When calculating the correlation coefficient between $\gamma$ and $\langle{T^{(1)}}\rangle$ on each network, the associated correlation coefficients are larger than 0.93, which indirectly demonstrates that the growth rate $\gamma$ in Eq.~(\ref{e6}) is closely related to the mean first passage time $\langle{T^{(1)}}\rangle$.

\section{The logarithmic growth pattern of maximal entropy random walks}\label{sec4}
\begin{figure}[!htb]
\centering
\includegraphics[width=0.95\textwidth,height=0.72\textheight]{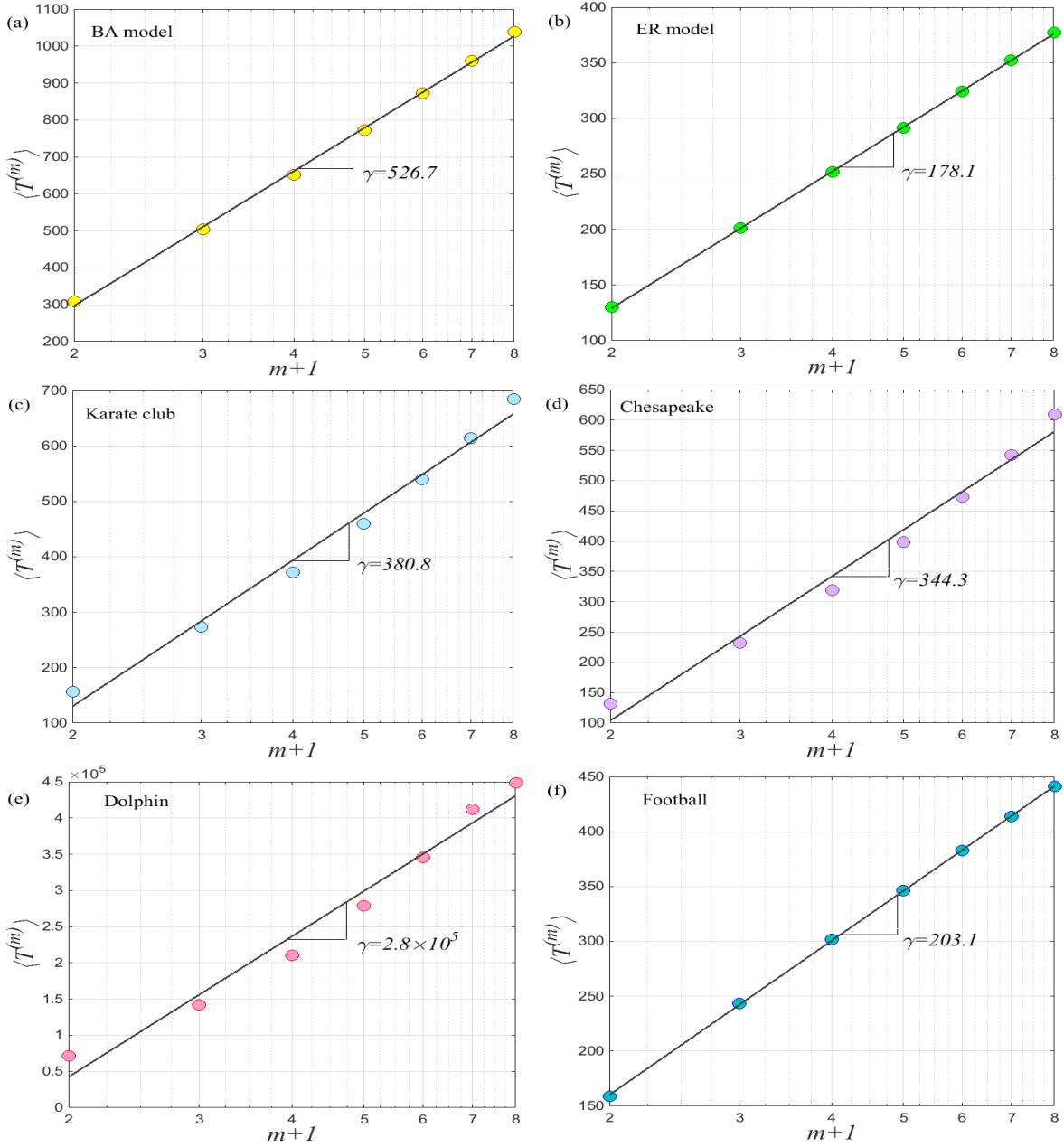}
\caption {(Color online) The semilogarithmic plots show the global MRCT $\langle{T^{(m)}}\rangle$ as a function of number of targets $m$ for maximal entropy random walks on (a) the BA model, (b) the ER model, (c) the ``Karate club'' network, (d) the ``Chesapeake'' network, (e) the ``Dolphin'' network, and (f) the ``Football'' network. The values of $\gamma$ are obtained from the slopes of the straight lines.} \label{fig4}
\end{figure}

We now study the problem of multi-target search based on the maximal entropy random walk strategy \cite{YLin2014}. The maximal entropy random walk is an unique biased diffusion process, where all trajectories of a given length and given endpoints are equiprobable. Such unusual property can lead to the Lifshitz phenomenon \cite{ZBurda2009} and has wide applications in network science, for example, detecting network community \cite{JKOchab2013}. The transition probability $p_{ij}$ of the maximal entropy random walk is
\begin{equation}
\ p_{ij}=\frac{a_{ij}}{\lambda}\frac{\mu_{j}}{{\mu_{i}}},
\label{e14}
\end{equation}
 where $\lambda$ is the largest eigenvalue of the adjacency matrix $A$ and $\mu_{i}$ is the $i\th$ element of the corresponding principal eigenvector $\mu$. Here, we investigate the global MRCT $\langle{T^{(m)}}\rangle$ as a function of number of targets $m$ for the maximal entropy random walks on different networks. Clearly, all profiles show the logarithmic growth behaviors of $\langle{T^{(m)}}\rangle$ vs $m$ as illustrated in Fig.~\ref{fig4}. Although the growth rate $\gamma$ changes significantly with respect to different networks, it is still highly related to the global MFPT $\langle{T^{(1)}}\rangle$, where the correlation coefficient between them is 0.95. These results provide further evidence that the logarithmic growth mechanism is a universal principle governing multi-target search. Moreover, we can now theoretically explain this interesting phenomenon using the annealed network approach  \cite{SNDorogovtsev2008} and the Sherman-Morrison formula \cite{JSherman1950}. We reinterpret its adjacency matrix $A$ as a weighted fully connected graph $\widetilde{A}$. In this situation, since the largest eigenvalue $\lambda=\frac{1}{N\langle{k}\rangle}\sum_{i}k_{i}^{2}$ and the corresponding eigenvector $\mu= (k_{1}/\sqrt{\sum_{i}k_{i}^2},k_{2}/\sqrt{\sum_{i}k_{i}^2},\cdots,k_{N}/\sqrt{\sum_{i}k_{i}^2})$ \cite{YLin2014}, the transition probability matrix $P$ of the maximal entropy random walks becomes
 \begin{equation}
\ P=\frac{1}{\sum_{i=1}^{N}k_{i}^{2}}\textbf{e}(k_{1}^{2},k_{2}^{2},\cdots,k_{N}^{2}),
\label{e15}
\end{equation}
which is clearly a special case of biased random walks with $\alpha=1$ given in Eq.~(\ref{e9}). Based on the theoretical result of biased random walks, the logarithmic growth pattern consequently establishes for multi-target search when adopting the maximal entropy random walk strategy. 

\section{Conclusions}\label{sec5}
In summary, we study random search processes for multi-target search on networks and provide an iterative method to determine the MRCT analytically, which links the gap between mean first passage time and cover time. Interestingly, we observe the emergence of the sublinear growth pattern occurring on multi-target search irrespective of the underlying network structure and random search strategy (i.e., generic random walks, biased random walks and maximal entropy random walks), which explores the generic growth mechanism of search time transiting from one single target (i.e., mean first passage time) to exhaustive searches (i.e., cover time). The sublinear growth mechanism reveals a universal law governing multiple target search. Moreover, our analysis also shows that for biased random walks, the global MRCT is minimized exactly when the global MFPT for a single target search is minimized, clearly exhibiting the robustness of the tuning parameter in the optimization of search processes. Our findings recover and extend the previous results shown in ref.~21, where the first moment of random cover time implies the logarithmic growth behavior of the search time versus the target number in the case of non-compact walks.

Moreover, in the process of deriving the MRCT, we only required that the stochastic motion satisfies the Markov property (i.e., memoryless) regardless of the exact form of the associated transition probability. Therefore, our analysis is applicable to a broad range of stochastic processes such as L\'{e}vy walks \cite{TFweng2015}, intermittent search strategies \cite{MChupeau2015}, and persistent random walks \cite{VTejedor2012}. In fact, our approach inherits and develops the traditional idea of Ref.~22, where it gives a fundamental formula for calculating mean first passage time. Meanwhile, we notice that mean first passage time is not always meaningful \cite{TGMattos2012,CMejiaMonasterio}, which hints the potential deficiency of using mean random cover time. In this situation, we may need to adopt other quantities instead of the MRCT for characterizing multi-target search. On the other hand, previous studies based on the Ref.~22 have seen that the eigenvalues and eigenvectors of an adjacency matrix associated with the network play a critical role in characterizing a single target search \cite{ZZZhang2012}. A more intriguing open problem is how to use the eigenvalues and eigenvectors of the adjacency matrix to describe and characterize multi-target search on networks. 

\begin{acknowledgments}
We thank Sara Alaee, Bahareh Harandizadeh, and Rui Zheng for valuable discussions and help. This research has been supported, in part, by General Research Fund 26211515 from the Research Grants Council of Hong Kong, and Innovation and Technology Fund ITS/369/14FP from the Hong Kong Innovation and Technology Commission. J.Z. is supported by the National Science Foundation of China (NSFC 61573107) and special Funds for Major State Basic Research Projects of China (2015CB856003).
\end{acknowledgments}

\appendix

\section{The relationship between mean random cover time and mean first passage time for two targets search}

We address how to express mean random cover time in terms of the associated mean first passage time for two targets search. Without loss of generality we assume that the two targets are placed at nodes $v_1$ and $v_2$. In this situation, if the first step of the walker is to node $v_{1}$ (resp. $v_{2}$), the expected number of steps required is $T_{v_{1},v_{2}}+1$ (resp. $T_{v_{2},v_{1}}+1$); if it is to some other node $j$, the expected number of steps becomes $T_{j,\{v_{1},v_{2}\}}+1$. Thus, for $i\neq{v_{1}}$ and ${v_{2}}$, we have
\begin{equation}
\ T_{i,\{v_{1},v_{2}\}}=p_{iv_{1}}(T_{v_{1},v_{2}}+1)+p_{iv_{2}}(T_{v_{2},v_{1}}+1)+\sum_{j\neq{v_{1},v_{2}}}p_{ij}(T_{j,\{v_{1},v_{2}\}}+1),
\label{s16}
\end{equation}
where $p_{iv_{1}}$ is the transition probability of the walker hopping from node $i$ to node $v_{1}$ at each time step. Since $T_{v_{1},\{v_{1},v_{2}\}}=T_{v_{1},v_{2}}$ and $T_{v_{2},\{v_{1},v_{2}\}}=T_{v_{2},v_{1}}$, Equation~(\ref{s16}) can be rewritten as
\begin{equation}
\ T_{i,\{v_{1},v_{2}\}}=1+\sum_{j}p_{ij}T_{j,\{v_{1},v_{2}\}}.
\label{s2}
\end{equation}
Similarly, let $r_{v_1,\{v_1,v_2\}}$ denote the expected number of steps required to revisit nodes $v_1$ and $v_2$ again starting from node $v_1$. In the same manner, $r_{v_1,\{v_1,v_2\}}$ can be represented as
\begin{equation}
\ r_{v_1,\{v_{1},v_{2}\}}=\sum_{j}p_{v_1j}(T_{j,\{v_{1},v_{2}\}}+1).
\label{s3}
\end{equation}
Combining Eq.~(\ref{s2}) and Eq.~(\ref{s3}) together, we obtain
\begin{equation}
\ (I-P)T^{(2)}=C-R,
\label{s4}
\end{equation}
where $I$ is the identity matrix, $C$ is an $N\times{\binom{N}{2}}$ matrix with all entries 1, and
\begin{equation*}
T^{(2)}=\left(\begin{array}{cccc}
 T_{1,\{1,2\}}&T_{1,\{1,3\}}& \cdots &T_{1,\{N-1,N\}}\\
T_{2,\{1,2\}}&T_{2,\{1,3\}}& \cdots &T_{2,\{N-1,N\}}\\
 \vdots&\vdots&\vdots&\vdots\\
T_{N,\{1,2\}}&T_{N,\{1,3\}}& \cdots &T_{N,\{N-1,N\}}
 \end{array} \right)_{N\times{\binom{N}{2}}},
 \end{equation*}
\begin{equation*}
R=\left(\begin{array}{cccc}
 r_{1,\{1,2\}}-T_{1,2}&r_{1,\{1,3\}}-T_{1,3}& \cdots &0\\
r_{2,\{1,2\}}-T_{2,1}&0&\cdots&0\\
 \vdots&\vdots&\vdots&\vdots\\
 0&0&\cdots&r_{N,\{N-1,N\}}-T_{N,N-1}
 \end{array} \right)_{N\times{\binom{N}{2}}},
 \end{equation*}
whose non-zero terms are the one for which the number of the line is one of the two elements of the tuple indexing the column.
Multiplying both sides of Eq.~(\ref{s4}) by the matrix W with each row being the stationary distribution vector $(w_{1},w_{2},\cdots,w_N)$, and using the fact that
\begin{equation}
\ W(I-P)=\textbf{0},
\label{s5}
\end{equation}
gives
\begin{equation}
\ w_{v_1}(r_{v_1,\{v_{1},v_{2}\}}-T_{v_{1},v_{2}})+w_{v_2}(r_{v_2,\{v_{1},v_{2}\}}-T_{v_{2},v_{1}})=1.
\label{s6}
\end{equation}
Since the matrix $(I-P+W)$ has an inverse \cite{Grinstead2006}, we denote $Z=(I-P+W)^{-1}$. In this situation, it is easy to verify that $ZC=C$ and $WZ=W$. Multiplying both sides of Eq.~(\ref{s4}) by $Z$ and using the evidence $(I-P+W)(I-W)=I-P$, we find the relation
\begin{equation}
\ T^{(2)}=C-ZR+WT^{(2)}.
\label{s7}
\end{equation}
From the above equation, we have
\begin{equation}
\ T_{i,\{v_1,v_2\}}=1-z_{i,v_1}(r_{v_1,\{v_1,v_2\}}-T_{v_1,v_2})-z_{i,v_2}(r_{v_2,\{v_1,v_2\}}-T_{v_2,v_1})+\sum_{j}w_{j}T_{j,\{v_1,v_2\}}.
\label{s8}
\end{equation}
Since $T_{v_{1},\{v_{1},v_{2}\}}=T_{v_{1},v_{2}}$ and $T_{v_{2},\{v_{1},v_{2}\}}=T_{v_{2},v_{1}}$, therefore
\begin{align}
\ T_{v_1,\{v_1,v_2\}}=1-z_{v_1,v_1}(r_{v_1,\{v_1,v_2\}}-T_{v_1,v_2})-z_{v_1,v_2}(r_{v_2,\{v_1,v_2\}}-T_{v_2,v_1})+\sum_{j}w_{j}T_{j,\{v_1,v_2\}}, 
\label{s9}\\
\ T_{v_2,\{v_1,v_2\}}=1-z_{v_2,v_1}(r_{v_1,\{v_1,v_2\}}-T_{v_1,v_2})-z_{v_2,v_2}(r_{v_2,\{v_1,v_2\}}-T_{v_2,v_1})+\sum_{j}w_{j}T_{j,\{v_1,v_2\}}.
\label{s10}
\end{align}
Combining Eq.~(\ref{s6}), Eq.~(\ref{s9}), and Eq.~(\ref{s10}) together, we obtain an explicit expression for $T_{i,\{v_1,v_2\}}$ as
\begin{equation}
\ T_{i,\{v_1,v_2\}}=\frac{T_{v_1,v_2}T_{v_2,v_1}+T_{i,v_1}T_{v_2,v_1}+T_{i,v_2}T_{v_1,v_2}}{T_{v_1,v_2}+T_{v_2,v_1}}.
\label{s13}
\end{equation}
The equation (\ref{s13}) shows the relation between mean random cover time and the associated mean first passage time for two targets search.

\bibliography{aipsamp}

\end{document}